\documentclass{ws-ijmpa}

\def\be{\begin{eqnarray}}
\def\en{\end{eqnarray}}

\def\ov{\overline}

\def\C{{\cal C}}
\def\T{{\cal T}}
\def\CP{{\it CP}~}

\begin{document}

\markboth{Hai-Yang Cheng}
 {Final-State Interactions in Hadronic $B$ Decays}

%
\catchline{}{}{}{}{}
%

\title{Effects of Final-State Interactions in Hadronic $B$ Decays}

\author{ Hai-Yang Cheng\index{Cheng, H. Y.} }

\address{Institute of Physics, Academia Sinica\\
Taipei, Taiwan 115, ROC\\
}

\maketitle

\begin{abstract}
Final-state rescattering effects on the hadronic $B$ decays and
their impact on direct and mixing-induced $CP$ asymmetries are
examined. Implications for some phenomenologies are briefly
discussed.

\end{abstract}

\section{Introduction}

Although the importance of final-state interactions (FSIs) has
long been recognized in hadronic charm decays, the general
folklore for hadronic $B$ decays is that FSIs are expected to play
only a minor role there due to the large energy release in the
energetic $B$ decay. However, there are growing hints at some
possible soft final-state rescattering effects in $B$ decays. The
measurement of the color-suppressed modes $B^0\to D^{0}\pi^0$,
when combined with the color-allowed $B\to D\pi$ decays, indicates
non-vanishing relative strong phases among various $B\to D\pi$
decay amplitudes. Denoting $\T$ and $\C$ as the color-allowed tree
amplitude and color-suppressed $W$-emission amplitude,
respectively, it is found that $\C/\T\sim 0.45\,{\rm exp}(\pm
i60^\circ)$ (see e.g. \cite{ChengBDpi}), showing a non-trivial
relative strong phase between $\C$ and $\T$ amplitudes. This is
the first evidence of large strong phases in charmful hadronic $B$
decays. Hence, it is natural to expect that sizable strong phases
can also manifest in charmless $B$ decays. Since the perturbative
strong phases are small in the conventional factorization
approach, it is likely that FSIs are responsible for the soft
strong phases.

One of the clear indications of large strong phases in charmless
hadronic $B$ decays arises from the recent measurements of direct
$CP$ violation. The first evidence of direct \CP violation was
reported by Belle \cite{Bellepipi} in $B^0\to\pi^+\pi^-$ even
before the $K^\pm\pi^\mp$ modes, but it has not been confirmed by
BaBar.\cite{BaBarpipi} A first confirmed observation of direct
$CP$ asymmetry was established last year in the charmless $B$
decays $\ov B^0(B^0)\to K^\mp\pi^\pm$ by both BaBar
\cite{BaBarKpi} and Belle. \cite{BelleKpi} Also the combined BaBar
and Belle measurements of $\ov B^0\to\rho^\pm\pi^\mp$ imply a
$3.6\sigma$ direct \CP asymmetry in the $\rho^+\pi^-$ mode.
\cite{HFAG}

Table 1 shows the comparison of the model predictions of direct
\CP asymmetries with the world averages of experimental results.
\cite{HFAG} The agreement of pQCD results \cite{pQCD} with
experiment for $K^-\pi^+$ and $\pi^+\pi^-$ is impressive,
recalling that pQCD predictions were made years before experiment
(for updated pQCD predictions, see \cite{LMS}).  In contrast, QCDF
predictions \cite{BBNS,BN} appear to be lousy as the predicted
signs for all the three modes $K^-\pi^+,\rho^+\pi^-,\pi^+\pi^-$
are wrong. This discrepancy has often led to the claims in the
literature that QCDF fails to describe direct $CP$ asymmetries.
However, this is not the case. As shown in Table I, there exist
several theoretical uncertainties in QCDF predictions especially
those arising from power corrections dictated by the last entry of
theoretical errors. Power corrections always involve endpoint
divergences. For example, the $1/m_b$ annihilation amplitude has
endpoint divergences even at twist-2 level and the hard spectator
scattering diagram at twist-3 order is power suppressed and posses
soft and collinear divergences arising from the soft spectator
quark. Since the treatment of endpoint divergences is model
dependent, subleading power corrections generally can be studied
only in a phenomenological way. While the endpoint divergence is
regulated in the pQCD approach by introducing the parton's
transverse momentum, \cite{pQCD} it is parameterized in QCD
factorization as
 \be
 X_A\equiv \int^1_0{dy\over y}={\rm ln}{m_B\over
 \Lambda_h}(1+\rho_A e^{i\phi_A}),
 \end{eqnarray}
with $\rho_A\leq 1$ and $\Lambda_h$ being a typical scale of order
500 MeV.

\begin{table}[h]
\tbl{Comparison of pQCD and QCD factorization (QCDF) predictions
of direct \CP asymmetries (in \%) with experiment. Also shown are
the QCDF results in scenario 4 denoted by QCDF(S4)$^{10}$ and the
FSI modifications to short-distance (SD) predictions.$^{11}$ The
pQCD results for $\rho\pi$ modes are taken from $^{12}$.}
{\begin{tabular}{| l r r r r r |}
\hline Mode & Expt. & pQCD & QCDF~~~~~~~~~~~~ & QCDF(S4) & SD+FSI  \\
 \hline
 $\ov B^0\to K^-\pi^+$ & $-11\pm2$ & $-17\pm5$  &
 $4.5^{+1.1+2.2+0.5+8.7}_{-1.1-2.5-0.6-9.5}$ & $-4.1$ & $-15^{+3}_{-1}$  \\
 $\ov B^0\to\rho^+\pi^-$ & $-47^{+13}_{-14}$ & $-7.1^{+0.1}_{-0.2}$  &
 $0.6^{+0.2+1.3+0.1+11.5}_{-0.1-1.6-0.1-11.7}$ & $-12.9$ & $-43\pm11$  \\
 $\ov B^0\to\pi^+\pi^-$ & $37\pm10$ & $23\pm7$  &
 $-6.5^{+2.1+3.0+0.1+13.2}_{-2.1-2.8-0.3-12.8}$ & 10.3 & $-$  \\
 $\ov B^0\to\pi^0\pi^0$ & $28\pm39$ & $30\pm10$ &
 $45.1^{+18.4+15.1+~4.3+46.5}_{-12.8-13.8-14.1-61.6}$ & $-19$ & $-30^{+1}_{-4}$ \\
 $\ov B^0\to \rho^-\pi^+$ & $-15\pm9$ & $12\pm2$ &
 $-1.5^{+0.4+1.2+0.2+8.5}_{-0.4-1.3-0.3-8.4}$ & 3.9 & $-24\pm6$ \\
 \hline
\end{tabular}}
\end{table}

Because of the large uncertainties in power corrections, one may
wonder if it is possible to accommodate the experimental
measurements of direct \CP violation in QCDF in certain parameter
space. This is indeed the case. Just like the pQCD approach where
the annihilation topology plays an essential role for producing
sizable strong phases and for explaining the penguin-dominated
$VP$ modes, Beneke and Neubert \cite{BN} chose a favorable
scenario (denoted as S4) to accommodate the observed
penguin-dominated $B\to PV$ decays and the measured sign of direct
\CP asymmetry in $\ov B^0\to K^-\pi^+$ by choosing $\rho_A=1$,
$\phi_A=-55^\circ$ for $PP$, $\phi_A=-20^\circ$ for $PV$ and
$\phi_A=-70^\circ$ for $VP$ modes. The sign of $\phi_A$ is chosen
so that the direct \CP violation $A_{K^-\pi^+}$ agrees with the
data. It is clear from Table I that the signs of $A_{\pi^+\pi^-}$
and $A_{\rho^+\pi^-}$ are correctly reproduced in QCD(S4). In
short, one needs large strong phases to explain the observed
direct \CP violation in $B$ decays.

For given $\rho_A$ and $\phi_A$, one can claim that QCDF still
makes predictions. However, the origin of these phases and large
annihilation magnitude is unknown. Moreover, the annihilation
topologies do not help enhance the $\pi^0\pi^0$ and $\rho^0\pi^0$
modes. Note that neither pQCD nor QCDF can explain the large
direct $CP$ asymmetry observed in $\ov B^0\to\rho^+\pi^-$. Hence,
one would wish to have an explanation of the data without invoking
weak annihilation. Therefore, it is of great importance to study
final-state rescattering effects on decay rates and \CP violation.

Besides the above-mentioned \CP violation, there exist several
other hints at large FSI effects in the $B$ sector. \cite{CCS} For
example, the measured branching ratio \cite{HFAG} ${\cal
B}(B^0\to\pi^0\pi^0)=(1.5\pm0.3)\times 10^{-6}$  cannot be
explained by either QCDF or pQCD and this may call for a possible
rescattering effect to induce $\pi^0\pi^0$. The QCDF predictions
for penguin-dominated modes such as $B\to
K^*\pi,~K\rho,~K\phi,~K^*\phi$ are consistently lower than the
data by a factor of 2 to 3. \cite{BN} This large discrepancy
between theory and experiment indicates the importance of
subleading power corrections such as the annihilation topology
and/or FSI effects.

Our goal is to study FSI effects on branching ratios and direct
\CP asymmetries in $B$ decays. Long distance (LD) rescattering
effects can be included in any SD approach but it requires
modelling of the $1/m_b$ power corrections. In \cite{CCS} we
present a specific model for FSI (to be described in the next
section) to predict (rather than accommodate) the sign and
magnitude of direct \CP violation.

\section{Final State Interactions in Hadronic $B$ decays}
In QCDF there are two hard strong phases: one from the absorptive
part of the penguin graph in $b\to s(d)$ transitions \cite{BSS}
and the other from the vertex corrections. However, these
perturbative strong phases do not lead to the correct sign of
direct \CP asymmetries observed in $K^-\pi^+$, $\rho^+\pi^-$ and
$\pi^+\pi^-$ modes. Therefore, one has to consider the
nonpertrubative strong phases induced from power suppressed
contributions such as FSIs. The idea is that if the intermediate
states are CKM more favored than the final state, e.g. charm
intermediate states in charmless $B$ decays, then the absorptive
part of the final-state rescattering amplitude can easily give
rise to large strong phases and make significant contributions to
the rates.

Based on the Regge approach, Donoghue {\it et al.} \cite{Donoghue}
have reached the interesting conclusion that FSIs do not disappear
even in the heavy quark limit and soft FSI phases are dominated by
inelastic scattering, contrary to the common wisdom. However, it
was later pointed out by Beneke {\it et al.} \cite{BBNS} within
the framework of QCD factorization that the above conclusion holds
only for individual rescattering amplitudes. When summing over all
possible intermediate states, there exist systematic cancellations
in the heavy quark limit so that the strong phases must vanish in
the limit of $m_b\to\infty$. Consequently, the FSI phase is
generally of order ${\cal O}(\alpha_s,\Lambda_{\rm QCD}/m_b)$. In
reality, because the $b$ quark mass is not very large and far from
being infinity, the aforementioned cancellation may not occur or
may not be very effective for the finite $B$ mass. Hence, the
strong phase arising from power corrections can be in principle
very sizable.

At the quark level, final-state rescattering can occur through
quark exchange and quark annihilation. In practice, it is
extremely difficult to calculate the FSI effects, but it may
become amenable at the hadron level where FSIs manifest as the
rescattering processes with $s$-channel resonances and one
particle exchange in the $t$-channel. In contrast to $D$ decays,
the $s$-channel resonant FSIs in $B$ decays is expected to be
suppressed relative to the rescattering effect arising from quark
exchange owing to the lack of the existence of resonances at
energies close to the $B$ meson mass. Therefore, we will model
FSIs as rescattering processes of some intermediate two-body
states with one particle exchange in the $t$-channel and compute
the absorptive part via the optical theorem. \cite{CCS}

The approach of modelling FSIs as soft rescattering processes of
intermediate two-body states has been criticized on several
grounds. \cite{BBNS} For example, there are many more intermediate
multi-body channels in $B$ decays and systematic cancellations
among them are predicted to occur in the heavy quark limit. This
effect of cancellation will be missed if only a few intermediate
states are taken into account. As mentioned before, the
cancellation may not occur or may not be very effective as the $B$
meson is not infinitely heavy. Hence, we may assume that two-body
$\rightleftharpoons$ $n$-body rescatterings are negligible either
justified from the $1/N_c$ argument or suppressed by large
cancellations. Indeed, it has been even conjectured that the
absorptive part of long-distance rescattering is dominated by
two-body intermediate states, while the dispersive part is
governed by multi-body states. \cite{Suzuki} At any rate, we view
our treatment of the two-body hadronic model for FSIs as a working
tool. We work out the consequences of this tool to see if it is
empirically working. If it turns out to be successful, then it
will imply the possible dominance of intermediate two-body
contributions.

The calculations of hadronic diagrams for FSIs involve many
theoretical uncertainties. Since the particle exchanged in the $t$
channel is off shell and since final state particles are hard,
form factors or cutoffs must be introduced to the strong vertices
to render the calculation meaningful in perturbation theory. We
shall parametrize the off-shell effect of the exchanged particle
as
 \be
 F(t,m)=\,\left({\Lambda^2-m^2\over \Lambda^2-t}\right)^n,
 \end{eqnarray}
normalized to unity at $t=m^2$ with $m$ being the mass of the
exchanged particle. The monopole behavior of the form factor (i.e.
$n=1$) is preferred as it is consistent with the QCD sum rule
expectation. \cite{Gortchakov} For the cutoff $\Lambda$, it should
be not far from the physical mass of the exchanged particle. To be
specific, we write $\Lambda=m_{\rm exc}+r\Lambda_{\rm QCD}$ where
the parameter $r$ is expected to be of order unity and it depends
not only on the exchanged particle but also on the external
particles involved in the strong-interaction vertex. As we do not
have first-principles calculations for form factors, we shall use
the measured decay rates to fix the unknown cutoff parameters and
then use them to predict direct \CP violation.

As mentioned in the Introduction, final-state rescattering effects
can be implemented in any SD approach. For our purpose, we shall
choose QCD factorization as the short-distance framework to start
with. Moreover, we should set $\rho_{A,H}$ to zero in order to
avoid the double counting problem.

\vskip 0.2cm \noindent {\bf 2.1~~ Penguin dominated modes} \vskip
0.2cm
 Penguin dominated modes such as $B\to K\pi,~K^*\pi,~K\rho,~\phi
K^{(*)}$ receive sizable contributions from rescattering of charm
intermediate states (i.e. the so-called long-distance charming
penguins). For example, the branching ratios of $B\to \phi K$ and
$\phi K^*$ can be enhanced from $\sim 5\times 10^{-6}$  predicted
by QCDF to the level of $1\times 10^{-5}$ by FSIs via rescattering
of charm intermediate states. \cite{CCS} The decay $\ov B^0\to
K^{*-}\pi^+$ predicted at the level $3.8\times 10^{-9}$ by QCDF is
enhanced by final-state rescattering to the order of $10\times
10^{-6}$, to be compared with $(12.7^{+1.8}_{-1.7})\times 10^{-6}$
experimentally. \cite{HFAG}

\vskip 0.2cm \noindent {\bf 2.2~~ Tree dominated modes}

The color-suppressed $\rho^0 \pi^0$ mode is slightly enhanced by
rescattering effects to the order of $1.3\times 10^{-6}$, which is
consistent with the weighted average $(1.9\pm0.6)\times 10^{-6}$
of the experimental values, $(1.4\pm0.7)\times 10^{-6}$ by BaBar
\cite{BaBarrho0pi0} and  $(3.12^{+0.88+0.60}_{-0.82-0.76})\times
10^{-6}$ by Belle. \cite{Bellerho0pi0} Note that the branching
ratio of $\rho^0\pi^0$ is predicted to be of order $0.2\times
10^{-6}$ in the pQCD approach, \cite{Lu} which is too small
compared to experiment as the annihilation contribution does
not help enhance its rate.\\

\vskip 0.2cm \noindent {\bf 2.3~~ Direct \CP asymmetries} \vskip
0.2cm The strong phases in charmless $B$ decays are governed by
final-state rescattering. We see from the last column of Table 1
that direct $CP$-violating partial rate asymmetries in $K^-\pi^+$
and $\rho^+\pi^-$ are significantly affected by final-state
rescattering and their signs are different from that predicted by
the short-distance QCDF approach. Direct \CP violation in
$\pi^+\pi^-$ cannot be predicted in this approach as charming
penguins are not adequate to explain the $\pi\pi$ data: the
predicted $\pi^+\pi^-$ ($\sim 9\times 10^{-6}$) is too large
whereas $\pi^0\pi^0$ ($\sim 0.4\times 10^{-6}$) is too small. This
means that a dispersive long-distance contribution is needed to
interfere destructively with $\pi^+\pi^-$ so that $\pi^+\pi^-$
will be suppressed while $\pi^0\pi^0$ will get enhanced. One needs
the observed $\pi\pi$ rates  and $A_{\pi^+\pi^-}$ to fix this new
LD contribution (see \cite{CCS} for details).

Direct \CP asymmetries in $\ov B^0\to\pi^0\pi^0,\rho^-\pi^+$
decays are also shown in Table 1 where we see that the predictions
of pQCD and the SD approach supplemented with FSIs are opposite in
sign. It will be interesting to measure direct \CP violation in
these two decays to test different models.

\section{Mixing-induced $CP$ violation}
Considerable activity in search of possible New Physics beyond the
Standard Model has recently been devoted to the measurements of
time-dependent \CP asymmetries in neutral $B$ meson decays into
final \CP eigenstates defined by
 \be
 {\Gamma(\ov B(t)\to f)-\Gamma(B(t)\to f)\over
 \Gamma(\ov B(t)\to f)+\Gamma(B(t)\to
 f)}=S_f\sin(\Delta mt)+A_f\cos(\Delta mt),
 \en
where $\Delta m$ is the mass difference of the two neutral $B$
eigenstates, $S_f$ monitors mixing-induced \CP asymmetry and $A_f$
measures direct \CP violation. The time-dependent {\it CP}
asymmetries in the $b\to sq\bar q$ penguin-induced two-body decays
such as $B^0\to (\phi,\omega,\pi^0,\eta',f_0)K_S$ and three-body
decays e.g. $B^0\to K^+K^-K^0_S,K_S^0K_S^0K_S^0$ measured by BaBar
\cite{BaBarS} and Belle \cite{BelleS} show some indications of
sizable deviations from the expectation of the SM where \CP
asymmetry in all above-mentioned modes should be equal to $\sin
2\beta$ inferred from the $B^0\to J/\psi K$ decay with a small
deviation {\it at most} ${\cal O}(0.1)$. \cite{LS} In order to
detect the signal of New Physics unambiguously in the penguin
$b\to s$ modes, it is of great importance to examine how much of
the deviation of $S_f$ from $S_{J/\psi K}$ is allowed in the SM.
Based on the framework of QCD factorization, the mixing-induced
\CP violation parameter $S_f$ in the seven 2-body modes
$(\phi,\omega,\rho^0,\eta',\eta,\pi^0,f_0)K_S$ has recently been
quantitatively studied in \cite{CCSsin2beta} and \cite{Beneke}. It
is found that the sign of $\Delta S_f\equiv -\eta_fS_f-S_{J/\psi
K_S}$ ($\eta_f$ being the \CP eigenvalue of the final state $f$)
at short distances is positive except for the channel $\rho^0K_S$.

In all previous studies and estimates of $\Delta S_f$, effects of
FSI were not taken into account. In view of the striking
observation of large direct \CP violation in $B^0\to
K^\pm\pi^\mp$, it is clear that final-state phases in two-body $B$
decays may not be small. It is therefore important to understand
their effects on $\Delta S_f$. It is found \cite{CCSsin2beta} that
the long-distance effects on $S_f$ are generally negligible except
for the $\omega K_S$ and $\rho^0K_S$ modes where $S_f$ is lowered
by around 15\% for the former and enhanced by the same percentage
for the latter and $\Delta S_{\omega K_S,\rho^0K_S}^{SD+LD}$
become consistent with zero within errors. Moreover, the central
values of $\Delta S_f$ become positive for all the modes under
consideration, but they tend to be rather small compared to the
theoretical uncertainties involved so that it is difficult to make
reliable statements on the sign at present. \cite{CCSsin2beta}

\begin{table}[h]
\tbl{Direct $CP$ asymmetry parameter $A_f$ and the mixing-induced
\CP parameter $\Delta S_f^{SD+LD}$ for various modes. The first
and second theoretical errors correspond to the SD and LD ones,
respectively (see $^{23}$ for details).}
{\begin{tabular}{l r c l r r c}\hline
 &  \multicolumn{3}{c}{$\Delta S_f$}
 &   \multicolumn{3}{c}{$A_f(\%)$}  \\ \cline{2-4} \cline{5-7}
\raisebox{2.0ex}[0cm][0cm]{State} & SD & SD+LD & Expt & SD & SD+LD
& Expt
\\ \hline
 $\phi K_S$ & $0.02^{+0.00}_{-0.04}$ & $0.03^{+0.01+0.01}_{-0.04-0.01}$ & $-0.38\pm0.20$  &
  $1.4^{+0.3}_{-0.5}$ & $-2.6^{+0.8+0.0}_{-1.0-0.4}$  & $4\pm17$ \\
 $\omega K_S$ & $0.12^{+0.05}_{-0.06}$  & $0.01^{+0.02+0.02}_{-0.04-0.01}$  & $-0.17^{+0.30}_{-0.32}$
 & $-7.3^{+3.5}_{-2.6}$  & $-13.2^{+3.9+1.4}_{-2.8-1.4}$ & $48\pm25$ \\
 $\rho^0K_S$ & $-0.09^{+0.03}_{-0.07}$ & $0.04^{+0.09+0.08}_{-0.10-0.11}$ &  & $9.0^{+2.2}_{-4.6}$ &
 $46.6^{+12.9+3.9}_{-13.7-2.6}$ &  \\
 $\eta' K_S$ & $0.01^{+0.00}_{-0.04}$ & $0.00^{+0.00+0.00}_{-0.04-0.00}$
 & $-0.30\pm0.11$  & $1.8^{+0.4}_{-0.4}$ & $2.1^{+0.5+0.1}_{-0.2-0.1}$ & $4\pm8$ \\
 $\eta K_S$ & $0.07^{+0.02}_{-0.04}$ & $0.07^{+0.02+0.00}_{-0.05-0.00}$ &
 & $-6.1^{+5.1}_{-2.0}$ & $-3.7^{+4.4+1.4}_{-1.8-2.4}$ &   \\
 $\pi^0K_S$ & $0.06^{+0.02}_{-0.04}$ & $0.04^{+0.02+0.01}_{-0.03-0.01}$ & $-0.39^{+0.27}_{-0.29}$ &
 $-3.4^{+2.1}_{-1.1}$ & $3.7^{+3.1+1.0}_{-1.7-0.4}$ & $-8\pm14$  \\
 \hline
\end{tabular}}
\end{table}

Recently, we have also studied the decay rates and time-dependent
\CP asymmetries in the three-body decays $B^0\to K^+K^-K_{S(L)}$
and $K_S K_S K_{S(L)}$ within the framework of factorization.
\cite{CCS3K} Owing to the presence of color-allowed tree
contributions in $B^0\to K^+K^-K_{S(L)}$, this penguin-dominated
mode is subject to a significant tree pollution and the deviation
of the mixing-induced \CP asymmetry from that measured in $B\to
J/\psi K_S$, namely, $\Delta \sin 2\beta_{K^+K^-K_{S(L)}}\equiv
\sin 2\beta_{K^+K^-K_{S(L)}}-\sin 2 \beta_{J/\psi K_S}$, can be as
large as ${\cal O}(0.10)$. The deviation $\Delta
\sin2\beta_{K^+K^-K_{S(L)}}$ arises mainly from the large
$m_{K^+K^-}$ region.

\section{Polarization Anomaly in $B\to\phi K^*$}
For $B\to V_1V_2$ decays with $V$ being a light vector meson, it
is expected that they are dominated by longitudinal polarization
states and respect the scaling law: $1-f_L={\cal O}(m_V^2/m_B^2)$.
However, a low value of the longitudinal fraction $f_L\approx
50\%$ and sizable perpendicular polarization $f_\bot\approx 20\%$
in $\phi K^*$ decays were observed by both BaBar \cite{BaBarVV}
and Belle \cite{BelleVV} (see Table 3). The polarization anomaly
for $f_L$ poses an interesting challenge for any theoretical
interpretation.

\begin{table}[h]   \label{tab:BphiKV}
 \tbl{ Experimental data for
\CP averaged branching ratios (in units of $10^{-6}$) and
polarization fractions for $B\to\phi K^*$ and $\rho K^*$.
$^{27,28}$} {\begin{tabular}{c c c  c} \hline
Mode & BaBar & Belle & Average  \\
\hline
 $f_L(\phi K^{*0})$
              & $0.52\pm0.05\pm0.02$
              & $0.45\pm0.05\pm0.02$
              & $0.48\pm0.04$
              \\
 $f_\bot(\phi K^{*0})$
              & $0.22\pm0.05\pm0.02$
              & $0.31^{+0.06}_{-0.05}\pm0.02$
              & $0.26\pm0.04$
              \\
 $f_L(\phi K^{*+})$
              & $0.46\pm0.12\pm0.03$
              & $0.52\pm0.08\pm0.03$
              & $0.50\pm0.07$
              \\
 $f_\bot(\phi K^{*+})$
              &
              & $0.19^{+0.08}_{-0.02}$
              & $0.19\pm0.08$
              \\
 \hline
 $f_L(\rho^0 K^{*+})$
              & $0.96^{+0.04}_{-0.15}\pm0.04$
              &
              & $0.96^{+0.06}_{-0.15}$
              \\
 $f_L(\rho^+ K^{*0})$
              & $0.79\pm0.08\pm0.04$
              & $0.43\pm0.11^{+0.05}_{-0.02}$
              & $0.66\pm0.07$
              \\
 \hline
\end{tabular}}
\end{table}

Working in the context of QCD factorization, it has been argued
that the lower value of the longitudinal polarization fraction and
the large transverse rate can be ``accommodated'' by the
$(S-P)(S+P)$ penguin-induced annihilation contributions.
\cite{Kagan} This is so because the transverse polarization
amplitude induced from the above annihilation topologies is of the
same $1/m_b$ order as the longitudinal one.  For other alternative
solutions to the $\phi K^*$ anomaly, see \cite{Hou04}.

Contrary to the SD approach, it is considerably easy to obtain a
large transverse polarization via final state rescattering. To
illustrate the idea, consider the long-distance rescattering
processes from the intermediate states $D^{(*)}D_s^{(*)}$
\cite{CCS,Colangelo}. It is easy to find out the polarization
states of $D^*D_s^*$ in $B\to D^*D_s^*$ to be $f_L\approx 0.51$,
$f_\parallel\approx 0.41$ and $f_\bot\approx 0.08$. Hence, the
large transverse polarization induced from $B\to D^*D_s^*$ can be
propagated to $\phi K^*$ via FSI rescattering. What about $f_\bot$
? It is obvious that the perpendicular polarization induced from
$D^{(*)}D_s^{(*)}$ through rescattering is too small to account
for experiment. Indeed, $f_\bot$ vanishes in $m_c/m_b\to 0$ limit.
Nevertheless, rescattering from $B\to D^*D_s$ or $B\to DD^*_s$
have unique contributions to the $A_\bot$ amplitude.

However, we found no sizable perpendicular polarization owing
mainly to the large cancellations occurring in the processes $B\to
D_s^* D\to\phi K^*$ and $B\to D_s D^*\to\phi K^*$ and this can be
understood as a consequence of \CP and SU(3) symmetry. \cite{CCS}
Our result is thus drastically different from a recent similar
study in \cite{Colangelo}. In short, it is ``trivial" to get a
large $\phi K^*$ transverse polarization via LD rescattering, but
it takes some efforts e.g. the $p$-wave charm intermediate states
\cite{CCS} to circumvent the aforementioned CPS constraint and
achieve sizable perpendicular polarization.

As for the transverse polarization in $B\to\rho K^*$ decays, both
final-state rescattering and large annihilation scenarios lead to
$f_L(\rho K^*)\sim 60\%$. However, none of the existing models can
explain the observed disparity between $f_L(\rho^+ K^{*0})$ and
$f_L(\rho^0 K^{*+})$. This should be clarified both experimentally
and theoretically.

\section{Acknowledgments}

I am grateful to Chun-Khiang Chua and Amarjit Soni for very
fruitful collaboration and to Kuang-Ta Chao and Xiangdong Ji for
organizing this wonderful conference.

\end{document}